\documentclass[journal]{IEEEtran}
\usepackage{amsmath,amsfonts}
\usepackage{nicefrac}
\usepackage{amsmath,amsfonts,amssymb,amsthm}

\usepackage{bm}
\usepackage{setspace}
\usepackage[colorlinks,urlcolor=blue,linkcolor=blue,citecolor=blue]{hyperref}
\usepackage{lineno}
\usepackage{wrapfig}
\usepackage{color,array}
\usepackage{mathtools}
\usepackage{makecell}
\usepackage{multirow}
\usepackage[ruled, linesnumbered]{algorithm2e}
\usepackage{booktabs}
\usepackage[caption=false,font=normalsize]{subfig}
\usepackage{textcomp}
\usepackage{stfloats}
\usepackage{url}
\usepackage{verbatim}
\usepackage{graphicx}
\usepackage[table,xcdraw]{xcolor}
\usepackage{cite}
\usepackage{balance}
\usepackage[capitalize]{cleveref}

\newcommand{\mycommentstyle}[1]{\textcolor[HTML]{0565A6}{\small{#1}}}
\SetKwComment{Comment}{\mycommentstyle{// }}{}

\newcommand{\method}{CPIG\xspace}

\DeclareMathOperator*{\argmin}{\arg\!\min}

\begin{document}

\title{\method: Leveraging Consistency Policy with \\Intention Guidance for Multi-agent Exploration}

\author{
Yuqian~Fu,~\IEEEmembership{Graduate Student Member,~IEEE,} Yuanheng~Zhu,~\IEEEmembership{Senior Member,~IEEE,} Haoran~Li,~\IEEEmembership{ Member,~IEEE,} Zijie~Zhao,~\IEEEmembership{Graduate Student Member,~IEEE,} Jiajun~Chai,~\IEEEmembership{Graduate Student Member,~IEEE,} and~Dongbin~Zhao,~\IEEEmembership{Fellow,~IEEE}
% Anonymous Author(s)

\thanks{
This work has been submitted to the IEEE for possible publication. Copyright may be transferred without notice, after which this version may no longer be accessible.

The authors are with the State Key Laboratory of Multimodal Artificial Intelligence Systems, Institute of Automation, Chinese Academy of Sciences, Beijing 100190, China, and also with the School of Artificial Intelligence, University of Chinese Academy of Sciences, Beijing 100049, China.}
}

\markboth{Journal of \LaTeX Class Files, Vol. 1, No. 1, November 2024}
{First A. Author \MakeLowercase{\textit{et al.}}}

\maketitle

\begin{abstract}
    Efficient exploration is crucial in cooperative multi-agent reinforcement learning (MARL), especially in sparse-reward settings.
	However, due to the reliance on the unimodal policy, existing methods are prone to falling into the local optima, hindering the effective exploration of better policies.
    Furthermore, in sparse-reward settings, each agent tends to receive a scarce reward, which poses significant challenges to inter-agent cooperation.
    This not only increases the difficulty of policy learning but also degrades the overall performance of multi-agent tasks.
	To address these issues, we propose a Consistency Policy with Intention Guidance (\method), with two primary components:
	(a) introducing a multimodal policy to enhance the agent's exploration capability,
	and (b) sharing the intention among agents to foster agent cooperation.
	For component (a), \method incorporates a Consistency model as the policy, leveraging its multimodal nature and stochastic characteristics to facilitate exploration.
	Regarding component (b), we introduce an Intention Learner to deduce the intention on the global state from each agent's local observation.
	This intention then serves as a guidance for the Consistency Policy, promoting cooperation among agents.
	The proposed method is evaluated in multi-agent particle environments (MPE) and multi-agent MuJoCo (MAMuJoCo).
    Empirical results demonstrate that our method not only achieves comparable performance to various baselines in dense-reward environments but also significantly enhances performance in sparse-reward settings, outperforming state-of-the-art (SOTA) algorithms by 20\%.
\end{abstract}


\section{Introduction}
\IEEEPARstart{R}{ecent} years have witnessed a growing body of applications in cooperative multi-agent reinforcement learning (MARL), such as multi-robot tasks~\cite{hu2023neuronsmae} and autonomous driving~\cite{zhang2022trajgen}. Despite these successful applications, cooperative MARL still faces challenges in exploration due to limitations in the policy class regarding multi-modality and the necessity for cooperation in multi-agent systems (MAS) \cite{rizk2018decision,hao2023exploration}.

\begin{figure}
	\centering	\includegraphics[width=\linewidth]{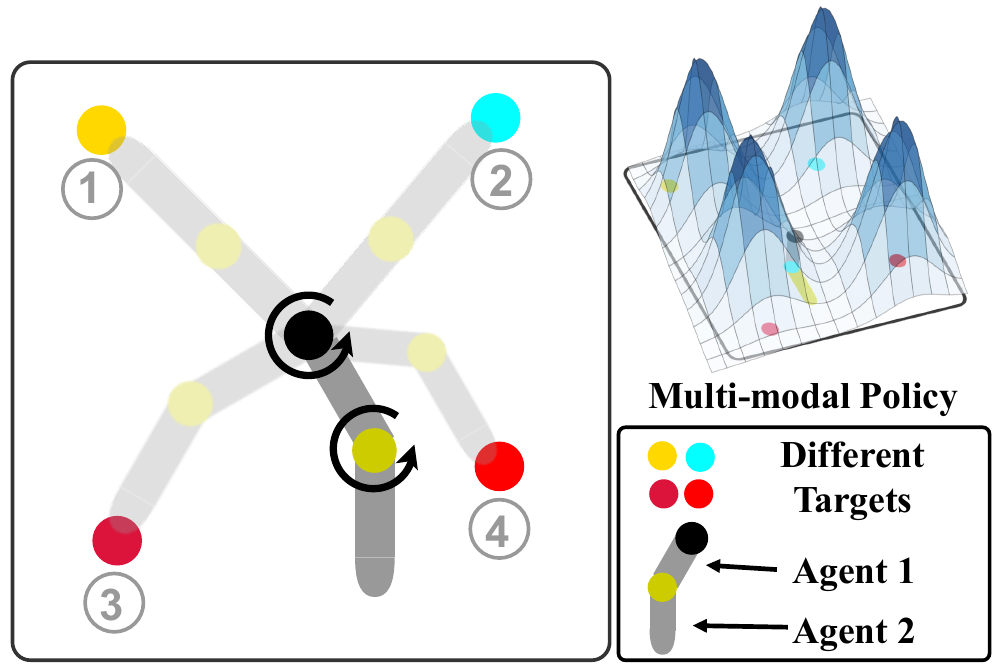}
	\caption{
	\textbf{An example of cooperative exploration. }
The two-agent arm requires collaborative exploration to reach four targets at different locations.
In a sparse reward setting, agents should reach all targets before receiving any reward, making exploration and cooperation more challenging.
The top-right corner illustrates a visualization of the multimodal joint policy.
	}
	\label{fig:toyexm}
\end{figure}

In single-agent RL, the exploration quality heavily depends on the chosen policy class of agents~\cite{mazoure2020leveraging, huang2023reparameterized,sridhar2023nomad}, with inappropriate policy classes potentially leading to local optima.
This issue becomes more pronounced in MARL owing to the complex interaction between agents.
In MARL, popular methods~\cite{lowe2017multi,kuba2022trust,yu2022surprising,yin2024deep} commonly formulate the continuous policy as a unimodal density function, typically a Gaussian distribution.
While computationally efficient, these policies can significantly weaken the exploration, as the sampled actions tend to be concentrated around one modality.
Besides, the unimodal policy is prone to converging towards a suboptimal policy due to the lack of expressiveness, overfitting the behavior of other agents.
To address these challenges, some methods \cite{kumar2019stabilizing,mazoure2020leveraging,ren2021probabilistic} explore alternative policy classes to enhance exploration.
Although these methods improve exploration to some extent, they often exhibit limitations in practice.
For example, Gaussian mixture models can only cover a limited number of modes, and normalizing flow methods~\cite{kobyzev2020normalizing}, while able to compute density values, suffer from numerical instability due to their determinant dependence.
With the growing prevalence of diffusion models~\cite{song2021scorebased}, a series of works~\cite{wang2023diffusion,kang2023efficient,li2024stabilizing} apply them as a powerful and multi-modal policy class, primarily in the context of single-agent offline RL.
However, the diffusion model is time-intensive, involving multiple diffusion steps (\textit{e.g.}, 1000 steps), which makes the training and execution computationally expensive, especially for the online MARL heavily depends on sampling from environments.
In order to accelerate the sampling process, we employ a novel generative model, the consistency model, which is designed to map any diffusion point at any time step back to the start of the generative process based on the probability flow ordinary differential equation (PF-ODE)~\cite{song2021scorebased}.
In multi-agent systems, the cooperation among agents during exploration is critical in complex environments~\cite{wang2020influence,liu2021cooperative,hao2023exploration}.
To exemplify, consider the \texttt{Reacher4} task depicted in \cref{fig:toyexm}, wherein two agents are required to coordinate their swinging motions to ensure that the end-effector of the manipulator contacts one of the four targets.
In this scenario, the agents need to coordinate their actions to set the entire robotic arm in motion, instead of engaging in independent and meaningless actions.
In a sparse reward setting, agents must reach \textit{all} targets before receiving any reward, making exploration more challenging.
Additionally, as illustrated in the top-right corner of the example, we visualize how this environment requires policies to handle a multimodal distribution.
The agents must adapt to the multimodality introduced by the four targets, avoiding premature convergence to a unimodal policy and enabling simultaneous exploration of multiple targets.
This challenge can be tackled through the use of communication-based MARL techniques~\cite{hu2021event}.
However, these techniques introduce challenges such as selecting appropriate information to transmit and additional bandwidth requirements.

Recently, some studies~\cite{xu2023consensus,ruan2023learning,wang2024long} leverage a shared feature among agents, such as intention or consensus, to promote cooperation.
This concept stems from the fact that \emph{at each timestep, while the local observations of individual agents are unique, they collectively represent different facets of the same global state.}
In the given scenario, though observations of the agents only cover their local information such as angular velocity, this information also serves as a partial representation of the global state.
Therefore, these individual observations can be viewed as projections of the global state.
In this paper, one of our primary objectives is to extract the state information through an intention learner, thereby fostering cooperation.

Building on the aforementioned observations, we propose a novel framework for MARL, termed Consistency Policy with Intention Guidance (\method), designed to facilitate efficient cooperative exploration.
In this framework, we adopt the consistency model as the policy class, leveraging its powerful expressiveness and stochastic nature to facilitate multimodal exploration.
Compared to diffusion models, consistency models enable efficient one-step generation while preserving the benefits of multi-step iterative sampling.
To enhance cooperative exploration, we incorporate discrete intention as guidance for the consistency policy.
Specifically, we employ a codebook for a discrete, distinguishable intention representation.
This allows agents to derive the same estimation of the global state from a unified intention codebook, thereby guiding cooperative behaviors.
Besides, to strike a balance between exploration and exploitation during training, some studies \cite{sridhar2023nomad,zangirolami2024dealing} use the mask to decide on exploration or exploitation.
Building on insights from these prior works, we propose a guidance mask that periodically drops the guidance with a certain probability.
When guidance is dropped, agents tend to engage in individual random exploration. In contrast, with guidance, their behavior shifts toward a focus on exploitation and cooperative behaviors.
Additionally, during the initial phase of training, generative policies may produce unreliable actions, which can affect the training of the policy.
To address this, we introduce a self-reference mechanism during the training phase, leveraging past successful experiences to constrain the actions generated by the policy, thus facilitating exploration.

We conduct empirical evaluations of the proposed method on two distinct environments: the multiple-particle environment (MPE)~\cite{lowe2017multi} and multi-agent MuJoCo (MAMuJoCo)~\cite{peng2021facmac}.
Across all experiments, we explore both dense-reward settings and sparse-reward settings.
The sparse-reward settings present significant challenges for exploration, as agents receive rewards only upon completing a specified task.
The results demonstrate that \method exhibits notable superiority over competitive baselines in sparse-reward settings and achieves comparable performance in dense-reward settings, thereby underscoring the effectiveness of our approach.

Our main contributions are summarized as follows:
\begin{enumerate}
    \item We propose a Consistency Policy for agents, which can complete the diffusion process in a single step, enabling them to explore the environment in a multimodal manner.
    To the best of our knowledge, our method is the first to leverage consistency policies in MARL.
	\item We propose an Intention Learner to infer the global intention from local observations, aimed at guiding efficient cooperative exploration by sharing the same intention between agents.
    \item We incorporate a Self-reference Mechanism to constrain the generated actions by leveraging past successful experiences, thereby reducing the probability of generating invalid actions by policies.
    \item We evaluate our method on MPE and MAMuJoCo with dense and sparse reward settings.
    The results indicate that \method performs comparably to baselines in dense reward environments, while in sparse reward environments, which are more challenging for exploration, our algorithm outperforms SOTA algorithms by 20\%.
\end{enumerate}

\section{Related Work}
\subsection{Exploration in MARL}
As one of the most critical issues in reinforcement learning, the exploration has drawn great attention in single-agent RL research.
This can be even more challenging in complex environments
with sparse reward~\cite{hao2023exploration}.
In the multi-agent domain, several works \cite{mahajan2019maven,wang2020influence,liu2021cooperative} have emerged to address exploration challenges.
MAVEN \cite{mahajan2019maven}, for instance, aims to maximize the mutual information between the trajectories and latent variables, thereby enabling the learning of diverse exploration policies.
EITI and EDTI~\cite{wang2020influence}, on the other hand, are designed to capture the influence of one agent's behaviors on others, encouraging agents to visit states that influence the behavior of others.
CMAE~\cite{liu2021cooperative} adopts a shared exploration goal derived from multiple projected state spaces.
Beyond these considerations, recent works highlight that the choice of policy class employed by the agents can significantly impact exploration dynamics~\cite{mazoure2020leveraging, huang2023reparameterized}.
However, these efforts primarily concentrate on single-agent domains. 
In this paper, we highlight the equally crucial role of policy classes in decision-making within multi-agent systems.
In our proposed approach, \method, we introduce an innovative policy class, the intention-guided consistency policy, specifically tailored to enhance exploration in MARL.

\subsection{Diffusion Models for Decision Making}
The application of diffusion models in single-agent RL has emerged as a significant trend, showcasing their capability to enhance model expressiveness and improve the decision-making process~\cite{zhu2023diffusion}.
Recognized for their powerful and flexible representation, diffusion models find utility in various domains such as trajectory generation \cite{ajay2023is}
and latent skill extraction \cite{venkatraman2023reasoning}. 
Moreover, owing to their exceptional capacity for representing multimodal distributions, diffusion models are also employed as effective policy classes. 
In the realm of offline RL, methods like Diffusion-QL \cite{wang2023diffusion} and EDP \cite{kang2023efficient} replace conventional Gaussian policies with diffusion models.
Building upon the foundational work laid by Janner~\textit{et al.}~\cite{janner2022planning}, and Wang~\textit{et al.}~\cite{wang2020influence}, recent studies \cite{zhu2023madiff,li2023beyond} extend diffusion models to MARL, applying them to trajectory generation and policy estimation.
However, it is worth noting that diffusion models typically entail multiple sampling steps, posing a considerable time constraint for diffusion policy methods, particularly in online environments.
This challenge is further exacerbated in MARL due to the heightened computational demands as the number of agents increases~\cite{chai2023nvif,chai2024aligning}.
To address this issue, some methods adopt faster sampling techniques such as 
DPM-Solver~\cite{lu2022dpmsolver}.
Additionally, certain methods like CPQL~\cite{chen2023boosting} and Consistency-AC~\cite{ding2023consistency} incorporate the consistency model~\cite{song2023consistency}, streamlining the sampling process within just one or two diffusion steps.
Inspired by these studies, we introduce consistency models as policies into MARL for the first time.
Overall, our method not only leverages the multi-modal nature of diffusion models but also significantly enhances the time efficiency.

\section{Preliminaries}

\subsection{Problem Formulation}
We consider a fully cooperative multi-agent task as a \textit{decentralized partially observable Markov decision process} (Dec-POMDP) \cite{oliehoek2016concise}, represented as a tuple $G=\left \langle \mathcal{S}, \mathcal{A}, \mathcal{U}, \mathcal{P}, r, \mathcal{O}, \Omega, \gamma \right \rangle$, where $\mathcal{A}$ represents the set of agents with $|\mathcal{A}|=n_a$.
At each time step, the environment generates the global state $s\in \mathcal{S}$, and each agent $a \in \mathcal{A}$ receives a unique local observation $o_a\in \mathcal{O}$, produced by the observation function $\Omega(s, a):\mathcal{S}\times\mathcal{A}\to \mathcal{O}$.
Subsequently, each agent $a$ selects and executes its own action $u_a\in \mathcal{U}$, resulting in a joint action $\bm{u}\in \mathcal{U}^{n_a}$ that induces a state transition according to the state transition function $\mathcal{P}(s,\bm{u}):\mathcal{S}\times \mathcal{U}^{n_a}\to \mathcal{S}$.
Meanwhile, the environment provides a global reward shared by all agents, determined by the reward function $r(s,\bm{u}):\mathcal{S}\times \mathcal{U}^{n_a} \to \mathbb{R}$.
$R=\sum_{t=0}^{T_e} \gamma^t r^t$ is the agent's total return, where ${T_e}$ denotes the time horizon, and $\gamma\in [0,1)$ represents the discount factor.
	$Q_a(o,u)=\mathbb{E}_{\{\bm{o}_{-a},\bm{u}_{-a}\}\sim \mathcal{B}}\left[R|o,u,\bm{o}_{-a},\bm{u}_{-a}\right]$ is the action value function for agent $a$, where $\bm{o}_{-a}$ and $\bm{u}_{-a}$ denotes the observations and actions of all agents except agent $a$.
   The experience replay buffer $\mathcal{B}$ contains the tuples $\left \langle \bm{o}, \bm{o'},\bm{u},r\right \rangle$.
	
	\subsection{Consistency Models}
	The diffusion model~\cite{song2021scorebased} addresses the multimodal distribution matching problem using a stochastic differential equation (SDE), while the consistency model \cite{song2023consistency} tackles a comparable probability flow ordinary differential equation (PF-ODE): 
${\mathrm{d} x_\tau}/{\mathrm{d} \tau}=-\tau \nabla \log p_\tau(x)$, where $\tau \in [0, T]$, $T>0$ is a fixed constant.
Here, $p_\tau(x)=p_{\text{data}}(x)\otimes \mathcal{N}(0,\tau^2 I)$ denotes the distribution of data $x_\tau$ at step $\tau$, where $\otimes$ denotes the convolution operation and $p_{\text{data}}(x)=p_0(x)$ represents the raw data distribution.
	The reverse process occurs along the solution trajectory $\{\hat{x}_\tau\}_{\tau\in[\epsilon, T]}$, with $\epsilon$ being a small constant close to $0$, employed for handling the numerical issues at the boundary.
	To accelerate the sampling process of a diffusion model, the consistency model reduces the required number of sampling steps to a significantly smaller value without substantially compromising model generation performance.
	Specifically, it approximates a parameterized consistency function $f_\theta: (x_\tau,\tau) \to x_\epsilon$, which is defined as a map from the noisy sample $x_\tau$ at step $\tau$ back to the original sample $x_\epsilon$, in contrast to the step-by-step denoising function $p_\theta(x_{\tau-1}|x_\tau)$ as the reverse diffusion process in diffusion model.
	The consistency function possesses the property of self-consistency, meaning its outputs are consistent for arbitrary pairs of $(x_\tau,\tau)$ that belong to the same PF-ODE trajectory.
	Besides, for modeling the conditional distribution with condition variable $c$, the consistency function is modified to $f_\theta(c, x_\tau, \tau)$, representing a slight deviation from the original consistency model.
	
\section{Method}
\begin{figure*}[t]
	\centering	\includegraphics[width=\linewidth]{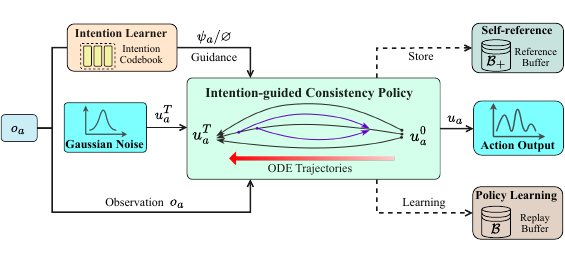}
	\caption{
		\textbf{The overall framework of the proposed \method.}
		Intention-guided diffusion policies facilitate cooperative exploration among multiple agents. 
The consistency policy utilizes the agent's observation and Gaussian noise as inputs, generating actions under the guidance of masked intention.
During the training phase, in addition to policy learning, the self-reference mechanism provides gradient back-propagation based on the disparity between the agent's action and those from the reference buffer, thereby imposing a policy constraint.}
	\label{fig:framework}
\end{figure*}

In this section, we introduce \method, a novel approach that integrates the consistency policy guided by the shared intention between agents for exploration.
As illustrated in \cref{fig:framework}, our framework consists of three principal components:
a) a consistency policy that generates actions from Gaussian noise,
b) a discrete intention codebook with masks, which serves as a guidance for cooperation,
and c) a self-reference mechanism to constrain consistency policies.
Detailed discussions on each component are provided in the subsequent sections.

\subsection{Consistency Policy}
\textbf{Notation.}
In this work, we distinguish between two types of timesteps, those associated with the diffusion process and those related to reinforcement learning. 
To clarify this distinction, we use superscripts $\tau \in[0, T]$ to represent diffusion timesteps and subscripts $t \in\{ 1, \dots, T_e \}$ to denote environment timesteps.

The consistency policy aims to explore in a multimodal way and avoid falling into local optima.
To facilitate understanding, we first formulate the diffusion policy before introducing the consistency policy, which serves as its successor.
In diffusion policy, the action of agent $a$ is diffused with a SDE:
\begin{equation}
	\label{eq:sde}
	\mathrm{d}u_a^\tau = \mu(u_a^\tau,o_a,\tau)\mathrm{d}\tau + \sigma(\tau) \mathrm{d}w^\tau,    
\end{equation}
where $\mu(\cdot,\cdot)$ representing the drift coefficient, $\sigma(\cdot)$ representing the diffusion coefficient, and $\{w^\tau\}_{\tau\in [0,T]}$ denoting the standard Brown Motion.
Beginning with $u_a^T$, the diffusion policy aims to recover the original action $u_a^0=u_a$ by solving a reverse process from $T$ to $0$ using the reverse-time SDE:
\begin{equation}
	\label{eq:re-sde}
	\mathrm{d}u_a^\tau = \left[\mu(u_a^\tau,o_a,\tau)-\sigma(\tau)^2\nabla \log p_\tau(u_a^\tau,o_a)\right]\mathrm{d}\tau+\sigma(\tau)\mathrm{d}\bar{w}^\tau,
\end{equation}
where $\bar{w}$ is the reverse Brown Motion, and the score function $\nabla \log p_\tau(u^\tau)$ is the sole unknown term at each diffusion timestep.
Solving the \cref{eq:re-sde} is challenging in practice, as an alternative, we solve the corresponding PF-ODE:
\begin{equation}
	\label{eq:pf-ode}
	\mathrm{d}u_a^\tau = \left[\mu(u_a^\tau,o_a,\tau)-\frac{1}{2}\sigma(\tau)^2 \nabla \log p_\tau(u_a^\tau,o_a)\right]\mathrm{d}\tau.
\end{equation}

Although this diffusion policy can be used for exploration, it is \emph{time-intensive} and makes the policy inference unacceptably slow, especially in \emph{tasks that contain multiple agents}.
To address this issue, we introduce a faster policy, consistency policy~\cite{song2023consistency,chen2023boosting}, based on \cref{eq:pf-ode}:
\begin{equation}
	\label{eq:con_po}
	\begin{aligned}
		\pi_{\theta_a}(u_a|o_a) & \coloneqq f_{\theta_a}(o_a,u_a^{\tau_n},{\tau_n})                                                      \\
		                        & =c_\mathrm{skip}({\tau_n})u_a^{\tau_n}+c_\mathrm{out}({\tau_n})F_{\theta_a} (o_a,u_a^{\tau_n},{\tau_n}), 
	\end{aligned}
\end{equation}
where the initial action $u_a^{\tau_n}\sim \mathcal{N}(0,{\tau_n} I)$, and $\theta_a$ represents the parameters of the policy $\pi$ of the agent $a$.
We discretize the diffusion horizon $[\epsilon, T]$ into a predetermined sequence $\{\tau_n|n\in N\}$ of length $N$ for determining the solution trajectory of action.
Moreover, $\tau_n$ is the $n$-th sub-intervals of the time horizon, with $n\sim \mathcal{U}(1,N-1)$.
The consistency function $F_{\theta_a}(o_a, u_a^\tau, \tau)$ is a trainable deep neural network that takes the observation $o_a$ of agent $a$ as input and outputs an action with the same dimensionality as $u_a^\tau$.
$c_{skip}$ and $c_{out}$ are differentiable functions, with $c_\mathrm{skip}(\epsilon)=1$, $c_\mathrm{out}(\epsilon)=0$, ensuring the consistency process remains differentiable at a fixed, small positive timestep $\epsilon$.

In this work, we employ the paradigm of centralized training with decentralized execution (CTDE).
Therefore, the joint policy is expressed as a composition of individual policies:
\begin{equation}
	\pi_{\bm{\theta}}(\bm{u}| \bm{o}) = \prod_{a=1}^{n_a} \pi_{\theta_a}(u_a | o_a).
\end{equation}

\textbf{Policy Learning.}
To train the consistency policy, we adopt an off-policy optimization approach similar to MADDPG \cite{lowe2017multi}.
Following the approach of Wang~\textit{et al.}~\cite{wang2023diffusion}, we integrate the Q-value function into the consistency model, training it to preferentially sample actions with higher values:
\begin{equation}
\label{eq:policy}
	\mathcal{L}_{\mathrm{policy}}(\theta_a)=-\mathbb{E}_{o_a\sim \mathcal{B},u_a\sim \pi_{\theta_a}(o_a)}\left[Q(o_a,u_a)\right].
\end{equation}

Specifically, as an estimation of the action value of the agent's current policy, the parameter $\beta$ of $Q(\cdot,\cdot)$ can be learned by minimizing the clipped double Q-learning loss \cite{fujimoto2018addressing}:
\begin{equation}
\label{eq:td}
	\begin{split}
		&\mathcal{L}_{\mathrm{TD}}(\beta_j )=\mathbb{E}_{\{\bm{o},\bm{u}_{-a}\}\sim \mathcal{B},u_a\sim \pi_{\theta_a}(o_a)}
		\\&\bigg[\Big(\big(r+\gamma \min_{i\in\{1,2\}}Q_{\beta_i^\intercal}(\bm{o}',{u_a}'\big)-Q_{\beta_j}\left(\bm{o},u_a\right)\Big)^2\bigg], j=1,2.
	\end{split}
\end{equation}
where $\beta,\beta^\intercal$ are the parameter and target parameter of Q-networks respectively. 

\subsection{Intention Guidance}

\begin{figure}[t]
	\centering
	\includegraphics[width=\linewidth]{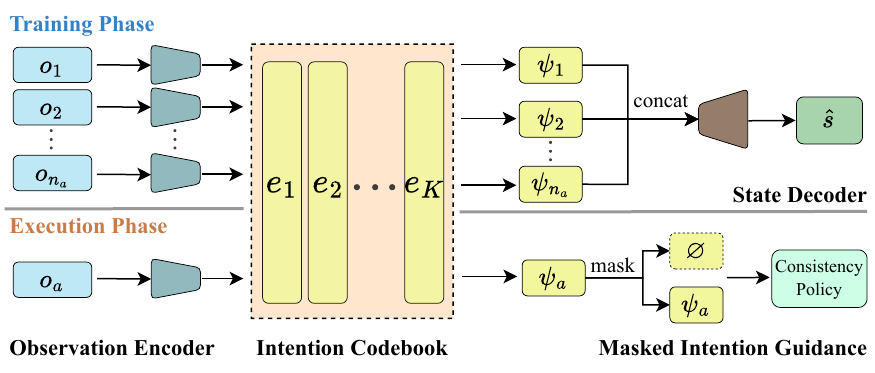}
	\caption{
	\textbf{The framework of the Intention Learner.} The intention learner consists of an observation encoder, an intention codebook, and a state decoder, which learns discrete intention representations by reconstructing states. It operates differently during the \textbf{\textcolor[HTML]{3382C7}{training}} and \textbf{\textcolor[HTML]{C77C49}{execution}} process.}
	\label{fig:intention}
\end{figure}

To guide cooperation in multi-agent exploration, we introduce an intention learner.
The intention represents an agent’s estimation of the global state, which influences its next action.
By sharing the same intention, multiple agents can exhibit coordinated behaviors.
We first formulate a set of $K$ discrete intentions $\Psi\coloneqq \{\psi_1,\dots,\psi_K\}$, where $K\in \mathbb{N}$ is a tunable hyperparameter.
Each intention $\psi_k$ is defined by a tuple $\left \langle e_{\psi_k},I_{\psi_k} \right \rangle$, where $k\in \{1,\dots,K\}$ is the identity of intention and $e_{\psi_k}\in \mathbb{R}^m$ is the code (or latent embedding) of intention $\psi_k$.
$I_{\psi_k}$ is the set of agents who have the same intention $\psi_k$, and each agent can only have one intention at each timestep: $I_{\psi_i} \cap I_{\psi_j} =\varnothing, \cup_j I_{\psi_j} =I$ for $i,j \in \{1,2,\dots, K\}$ and $i \neq j$.
To represent intention, we design an intention leaner with VQ-VAE~\cite{van2017neural}, as shown in \cref{fig:intention}. 
The motivations for this include:
a) The discrete codebook in VQ-VAE can effectively extract state features with similar semantics, providing better separation than traditional continuous representation methods.
b) The use of discrete intention representation helps mitigate the impact of unavoidable noise on exploration in stochastic environments.
c) The discretization process ensures a distinct association, where each agent is linked to a specific intention.

The intention learner comprises four primary components: an observation encoder, an intention codebook, a state decoder, and an intention mask.
While the intention can theoretically be derived from the global state $s$, the CTDE paradigm imposes restrictions, allowing each agent $a$ to access only its local observation $o_a$ during distributed execution.
Given that $o_a = \Omega(s, a)$ represents a projection of the global state $s$, we design a two-phase mechanism to infer intentions based on local observations.
To deduce the intention from local observations, the intention learner has \emph{two modes} during the training and execution phases:
During the \textbf{\textcolor[HTML]{3382C7}{training phase}}, the intention learner processes observations from all agents, obtains discrete intention representations for each agent's observation, and uses them to reconstruct the global state;
In the \textbf{\textcolor[HTML]{C77C49}{execution phase}}, it processes agent $a$'s observation to obtain the intention for guiding policy generation of action $u_a$.
In this manner, the intention learner develops the capacity to infer intention based on local observations.

\paragraph{Observation Encoder}
The encoder $z_e(\cdot;\alpha_e)$, parameterized by $\alpha_e$, encodes the observation of agent $o_a$ into an embedding $z_{e,a}\in \mathbb{R}^m$, matching the length of the intention representation $m$.
In practice, we use historical observations as inputs during each agent's execution to capture sufficient information about the global state.

\paragraph{Intention Codebook}
The process of inferring intention for each input observation proceeds as follows.
First, the embedding $z_{e,a}$ is mapped to a discrete index $k_a$ of the codebook $\mathcal{E}$:
\begin{equation}
\label{eq:intention}
	k_a = \argmin_{j\in\{1,\dots,K\}}\|z_{e,a}-e_j\|_2,
\end{equation}
where $e_j\in \mathcal{E}$ is a codebook vector, and $K$ denotes the size of the codebook.
Notably, the performance of intention learning is relatively insensitive to the choice of $K$, allowing us to set the number of discrete intentions in codebook $K=5$ across all scenarios.
Based on this discrete index $k_a$, the discrete intention representation for the agent $a$ is derived as $\psi_a=e_{k_a}$.
The loss function for the intention inference process is defined as:
\begin{equation}
\label{eq:vq}
	\|\mathrm{sg}(z_{e,a})-e_{k_a}\|_2^2+ \beta \|z_{e,a}-\mathrm{sg}(e_{k_a})\|_2^2,
\end{equation}
where $\mathrm{sg}$ denotes the straight-through gradient operation \cite{bengio2013estimating}, enabling back-propagation training of $\argmin$ operations, and $\beta$ is a hyperparameter balancing the degree of alignment between the codebook and input embeddings.
In practice, \method uses the exponential moving average (EMA) to update the codebook, instead of directly learning them as parameters.
Specifically, EMA updates the codebook $e_{k_a}$ by replacing it with a weighted combination of its previous value $e_{k_a}$ and the corresponding vector $z_{e, a}$:
 \begin{equation}
 \label{eq:ema}
     e_{k_a} \gets \mu z_{e,a}+(1-\mu)e_{k_a},
 \end{equation}
where the value of $\mu$ is a hyperparameter that controls the speed of the moving average updates.

\paragraph{State Decoder}
To develop the ability to deduce intention from local observations, we aggregate the intention $\psi_a$ from all agents together and process them through a state decoder $z_d(\cdot;\alpha_d)$, parameterized by $\alpha_d$, to reconstruct the global state $\hat{s}$.
We train the observation encoder and state decoder together by maximizing the log-likelihood of the reconstruction:
\begin{equation}
\label{eq:recon}
	\mathcal{L}_{\mathrm{recon}}(\alpha_{e,d})=\log p\left(\hat{s}|z_d\left(\bm{\psi};\alpha_d\right)\right),
\end{equation}
which encourages the encoding and decoding processes of the intention learner to capture the key information from partial observations for accurate intention deduction.
In addition, the reconstruction loss gradient is also passed to the encoder for training $\alpha_e$ by straight-through gradient estimation.

\paragraph{Mask Intention Exploration}
In this work, we employ two distinct exploration strategies: \textit{unguided exploration} and \textit{intention-guided exploration}, aiming to balance exploration and exploitation to some degree.
Unguided exploration allows agents to take action according to their current policies without specific intention guidance, emphasizing their self-directed exploration capabilities.
However, its effectiveness is limited by the initial policy quality and the inherent complexities of MAS, neglecting potential cooperative advantages.
Considering this, we introduce another exploration strategy, intention-guided exploration designed to steer the cooperation efforts of agents.
The intention-guided exploration instructs agents to take actions based on the consistency policy guided by their intention.
Through the guidance of shared intention among different agents, consistency policies can generate cooperative exploration behaviors.
Meanwhile, to incorporate intention guidance $\psi$, the consistency function $F_\theta(o,u,\tau)$ in \cref{eq:con_po} is modified to $F_\theta(o,u,\psi,\tau)$.

To implement the above concept and combine the two exploration mechanisms, we introduce a binary \emph{intention mask} $\mathcal{M}$.
This mask can modulate the influence of the intention vector $\psi_a$ on action generation by obscuring it through masked attention.
We achieve this through masked attention, setting the intention mask $\mathcal{M} = 0,\psi_a=\varnothing$ to prevent the downstream generation of $u_a$ from being guided by the intention representation.
Conversely, when $\mathcal{M} = 1$, the intention representation is incorporated alongside the agent’s observation during action generation.
During the \textbf{\textcolor[HTML]{3382C7}{training phase}}, the intention mask $\mathcal{M}$ is sampled from a Bernoulli distribution with a probability $p_{\mathcal{M}=1}$. 
We set a fixed value $p_{\mathcal{M}=1} =0.2$ during training, resulting in a higher proportion of training samples for intention-guided over unguided exploration.
In \textbf{\textcolor[HTML]{C77C49}{execution phase}}, we adjust $p_{\mathcal{M}=1}=1.0$, encouraging agents to prioritize intention-driven actions.
It's noteworthy that a recent work~\cite{chen2023boosting} demonstrates that consistency policies can achieve policy improvement with precise guidance.
In MARL, given the joint policy's complexity and the environment's non-stationarity, precise guidance can more significantly enhance the agents' policy improvement.

\subsection{Self-reference Mechanism}
Due to the nature of generative models, the consistency policy may generate invalid actions during the initial stages of training~\cite{chen2023generative}.
In order to constrain the generated actions, we propose a self-reference mechanism, inspired by self-imitation learning~\cite{oh2018self}.
The self-reference mechanism stores previous experiences in a replay buffer and learns to replicate actions that led to high returns in the past.
For each agent, we collect observation-action pairs $(o_a,u_a)$, along with their empirical return $R$, and store them in the replay buffer $\mathcal{B}$.
We then sample entries $(\bm{o}, \bm{u}, R)$ with empirical returns exceeding the action-value estimates $Q(\bm{o},u_a)$, creating a reference buffer $\mathcal{B}_+=\{(\bm{o}, \bm{u}, R)|R\geq Q(\bm{o},u_a)\}$, where $u_a$ is generated from the current policy $\pi_{\theta_a}$.
By uniformly sampling $(o_a, u_a)$ from $\mathcal{B}_+$, the consistency policy of agent $a$ can be trained using a reference loss function:
\begin{equation}
	\begin{split}
		\label{eq:ref}
		&\mathcal{L}_{\mathrm{ref}}(\theta_a)=\mathbb{E}_{n\sim \mathcal{U}(1,N-1),(o_a,u_a)\sim \mathcal{B}_{+}}
		\\&\left[\lambda(\tau_n)d\left(f_{\theta_a}(o_a,u^{\tau_{n+1}}_a,\psi_a,\tau_{n+1}),f_{\theta^\intercal_a}(o_a,u^{\tau_n}_a,\psi_a,\tau_n)\right)\right],
	\end{split}
\end{equation}
where $\lambda(\cdot)$ denotes a step-dependent weight function, $u^{\tau_n}=u+\tau_nz$ and $z\sim \mathcal{N}(0,I)$.
$d(\cdot,\cdot)$ represents the distance metric and we use $l_2$ distance $d(x, y) = \|x-y\|^2_2$.
Additionally, $f_{\theta^\intercal}$ is the exponential moving average of $f_\theta$, introduced for training stability.
The self-reference mechanism in our algorithm offers three benefits:
a) As generative methods are prone to generating invalid actions \cite{chen2023generative}, the self-reference mechanism acts as a policy constraint, ensuring more reliable action selection.
b) Learning from successful transitions can accelerate exploration during the early training stages.
c) self-imitation learning can be viewed as a lower bound of soft Q-learning~\cite{tang2020self}, promoting policy improvement.
The pseudocode of \method is illustrated in \cref{alg:cpcg}.

\begin{algorithm}[t]
\renewcommand{\arraystretch}{1.2}
\caption{Leveraging Consistency Policy with Intention Guidance for Multi-agent Exploration}
\label{alg:cpcg}
\Comment{\textcolor[HTML]{0671b9}{Initialize}}
Initialize the policy network $\pi_{\bm{\theta}}$ and critic networks $Q_{\beta_1},Q_{\beta_2}$\;
Initialize the replay buffer $\mathcal{B}$ and reference buffer $\mathcal{B}_+$\;
\For{episode $j=1,\dots,M$}{
Reset the environment and receive initial joint observation $\bm{o}$\;
\For{time step $t=1,\dots,H$}{
\Comment{\textcolor[HTML]{0671b9}{Intention guidance}}
For each agent $a$, infer intention $\psi_a$ by \cref{eq:intention}\;
Sample intention mask $\mathcal{M}=\{0,1\}$\;
Generate $u_a$ by $\pi_{\theta_a}$ with intention $\psi_a$, intention mask $\mathcal{M}$, and observation $o_a$\;
Update intention learner by \cref{eq:vq,eq:ema,eq:recon}\;
\Comment{\textcolor[HTML]{0671b9}{Buffer update}}
Execute actions $\bm{u}$ and observe reward $r$ and new observation $\bm{o}'$\;
Store $\left \langle \bm{o}, \bm{o'},\bm{u},r\right \rangle$ in replay buffer $\mathcal{B}$\;
Update reference buffer $\mathcal{B}_+$\;
\For{agent $a=1,\dots,n_a$}{
\Comment{\textcolor[HTML]{0671b9}{Policy update}}
Sample minibatch $B\in\mathcal{B}$ and $B_+\in\mathcal{B}_+$\;
Update policy $\pi_{\theta_a}$ by \cref{eq:policy}\; 
\Comment{\textcolor[HTML]{0671b9}{Self-reference}}
Constrain policy $\pi_{\theta_a}$ by self-reference mechanism \cref{eq:ref}\; 
}
\Comment{\textcolor[HTML]{0671b9}{Q-value update}}
Update Q-value networks $Q_{\beta_1},Q_{\beta_2}$ by \cref{eq:td}\;
}
}
\end{algorithm}
\section{Experiments}
In this section, we empirically evaluate our method to answer the following questions:
(1) Does \method effectively contribute to exploration and outperform baselines (See \cref{sec:results}, \cref{sec:visual})? 
(2) Can consistency policies alleviate the time-consuming issue associated with diffusion policies (See \cref{sec:time})? 
(3) How does the intention learner play a role in cooperative exploration (See \cref{sec:visual})? 
(4) Whether intention guidance and the self-reference mechanism contribute collectively to the final performance (See \cref{sec:abla})? 

\addtolength{\extrarowheight}{\belowrulesep}
\aboverulesep=0pt
\belowrulesep=0pt
\begin{table*}[ht]
  \renewcommand{\arraystretch}{1.2}
	\centering
	\caption{\textbf{Performance evaluation of returns and standard deviation in different
	scenarios with dense rewards and sparse rewards.} $\pm$ captures the standard deviation. We computed the average scores of each algorithm across different scenarios under two settings. \textbf{Highlighted} figures
	show the highest performance among each row.}
	\label{tab:comparison}
	\resizebox{\linewidth}{!}{%
	\begin{tabular}{>{}clrrrrrrr<{}}
		\toprule                               
		\multicolumn{2}{c}{\multirow{1}{*}{\textbf{Scenarios}}}                           & \multicolumn{1}{c}{\textbf{\method (Ours)}} & \multicolumn{1}{c}{\textbf{HATD3}~\cite{zhong2023heterogeneousagent}}  & \multicolumn{1}{c}{\textbf{MAPPO}~\cite{yu2022surprising}}  & \multicolumn{1}{c}{\textbf{HASAC}~\cite{zhong2023heterogeneousagent}}     & \multicolumn{1}{c}{\textbf{MAT}~\cite{wen2022multiagent}}       & \multicolumn{1}{c}{\textbf{CMAVEN}~\cite{mahajan2019maven}}
        & \multicolumn{1}{c}{\textbf{CMAE}~\cite{liu2021cooperative}} \\
		\midrule \multirow{6}{*}{\begin{tabular}[c]{@{}c@{}}\texttt{Dense}\\ \texttt{Reward}\end{tabular}}  & Navigation                                      & -68.1$\pm$2.0          & \textbf{-66.7$\pm$9.1} & -84.7$\pm$9.7             & -68.6$\pm$8.3             & -66.9$\pm$6.8         & -88.3$\pm$11.2 & -82.4$\pm$10.9 \\
		                                                                                  & Reference                                   & \textbf{-11.9$\pm$1.7} & -14.3$\pm$4.2          & -32.1$\pm$3.8             & -12.6$\pm$2.7             & -15.9$\pm$2.9         & -37.6$\pm$4.5 &  -21.1$\pm$3.9  \\
		                                                                                  & Reacher4 (2x1)                              & -22.1$\pm$0.4          & -22.3$\pm$0.8          & -22.9$\pm$0.5             & \textbf{-21.8$\pm$0.2}    & -23.0$\pm$0.2         & -26.9$\pm$0.3   & 23.4$\pm$0.6\\
		                                                                                  & HalfCheetah (2x3)                           & 6954.9$\pm$466.8       & 5651.3$\pm$441.3       & 5463.1$\pm$392.7          & \textbf{7541.3$\pm$472.6} & 6711.6$\pm$459.8      & 4526.8$\pm$314.2 
                                                                                          &5783.6$\pm$443.6\\
		                                                                                  & Hopper (3x1)                                & 3593.4$\pm$257.1       & 3349.9$\pm 187.6$      & 3228.3$\pm$194.9          & \textbf{3613.2$\pm$201.4} & 2745.8$\pm$174.1      & 2201.4$\pm$222.8 
                                                                                          & 2328.8$\pm$146.2\\	                                                                                  
& HalfCheetah (6x1)                               & \textbf{5692.7$\pm$339.4} & 4782.4$\pm$438.6          & 4083.1$\pm$352.1             & 5535.4$\pm$392.4             & 5057.4$\pm$334.7         & 3343.6$\pm$366.2    & 3253.0$\pm$401.7 \\\midrule		                                                                                  
\rowcolor[HTML]{F0F0FF} 
		\multicolumn{2}{c}{\multirow{1}{*}{\textbf{Average}}}       & 2689.8$\pm$177.9               & 2279.9$\pm$180.3       & 2105.8$\pm$158.9       & \textbf{2764.4$\pm$215.4} & 2401.4$\pm$195.7          & 1653.1$\pm$153.2       &1880.9$\pm$167.8\\
                                   
		\midrule \multirow{6}{*}{\begin{tabular}[c]{@{}c@{}}\texttt{Sparse}\\ \texttt{Reward}\end{tabular}} & Navigation                                      & \textbf{19.3$\pm$0.3}  & 8.9$\pm$1.3            & 4.2$\pm$0.8               & 10.4$\pm$0.6              & 3.9$\pm$0.9           & 4.9$\pm$1.8     & 7.6$\pm$1.1\\
		                                                                                  & Reference                                   & \textbf{18.1$\pm$0.5}  & 10.4$\pm$0.6           & 3.3$\pm$0.8               & 10.2$\pm$0.4              & 3.8$\pm$0.7           & 8.1$\pm$0.9   & 9.3$\pm$0.8  \\

& Reacher4 (2x1)                              & \textbf{50.0$\pm$0.0}  & 36.8$\pm$0.1           & 24.4$\pm$0.5              & 38.9$\pm$0.6              & 37.6$\pm$0.8          & 28.5$\pm$2.4 & 36.1$\pm$1.0   \\
		                                                                                  & HalfCheetah (2x3)                           & \textbf{965.0$\pm$8.0} & 805.0$\pm$12.6         & 703.7$\pm$27.9            & 801.2$\pm$10.9            & 733.7$\pm$14.1        & 629.1$\pm$13.2   &742.7$\pm$16.8\\
		                                                                                  & Hopper (3x1)                                & \textbf{-4.8$\pm$0.7}  & -33.6$\pm$3.8          & -27.9$\pm$2.7             & -16.9$\pm$0.9             & -20.3$\pm$4.3         & -18.4$\pm$3.3  &-21.9$\pm$2.3  \\

&  HalfCheetah (6x1)                                   & \textbf{704.3$\pm$9.8} & 501.6$\pm$10.2          & 533.6$\pm$21.9             & 622.0$\pm$16.4             & 578.1$\pm$18.2         & 480.7$\pm$10.3  & 541.5$\pm$17.2 \\\midrule
\rowcolor[HTML]{F0F0FF} 
		\multicolumn{2}{c}{\multirow{1}{*}{\textbf{Average}}}      & \textbf{291.9$\pm$3.8}         & 221.5$\pm$4.7          & 206.8$\pm$9.0          & 244.3$\pm$13.9            & 222.7$\pm$6.5             & 188.7$\pm$5.3   &  219.3$\pm$ 6.5    \\
		\bottomrule
	\end{tabular}
	}
\end{table*}

\subsection{Experimental Settings}

\paragraph{Multi-Agent Task Benchmark}
We conduct experiments in two widely-used multi-agent continuous control tasks including the Multi-agent Particle Environments\footnote{\url{https://pettingzoo.farama.org/environments/mpe}} (MPE)~\cite{lowe2017multi} and high-dimensional and challenging Multi-Agent MuJoCo\footnote{\url{https://github.com/schroederdewitt/multiagent_mujoco}} (MAMuJoCo)~\cite{peng2021facmac} tasks.
In MPE, agents known as physical particles need to cooperate to solve various tasks.
MAMuJoCo is an extension for MuJoCo locomotion tasks, originally designed for single-agent scenarios, enabling robots to move with the coordination of multiple agents.
Specifically, in MPE, we employ \texttt{Navigation} and \texttt{Reference} as the experimental environments.
In MAMuJoCo, the experimental environments include \texttt{HalfCheetah(2x3)}, \texttt{Hopper(3x1)}, \texttt{Reacher4(2x1)}, and \texttt{HalfCheetah(6x1)}.

\begin{figure*}[t]
    \centering
        \subfloat[Hopper (3x1)]{
        \begin{minipage}{0.19\linewidth}
            \centering 
            \includegraphics[width=\linewidth]{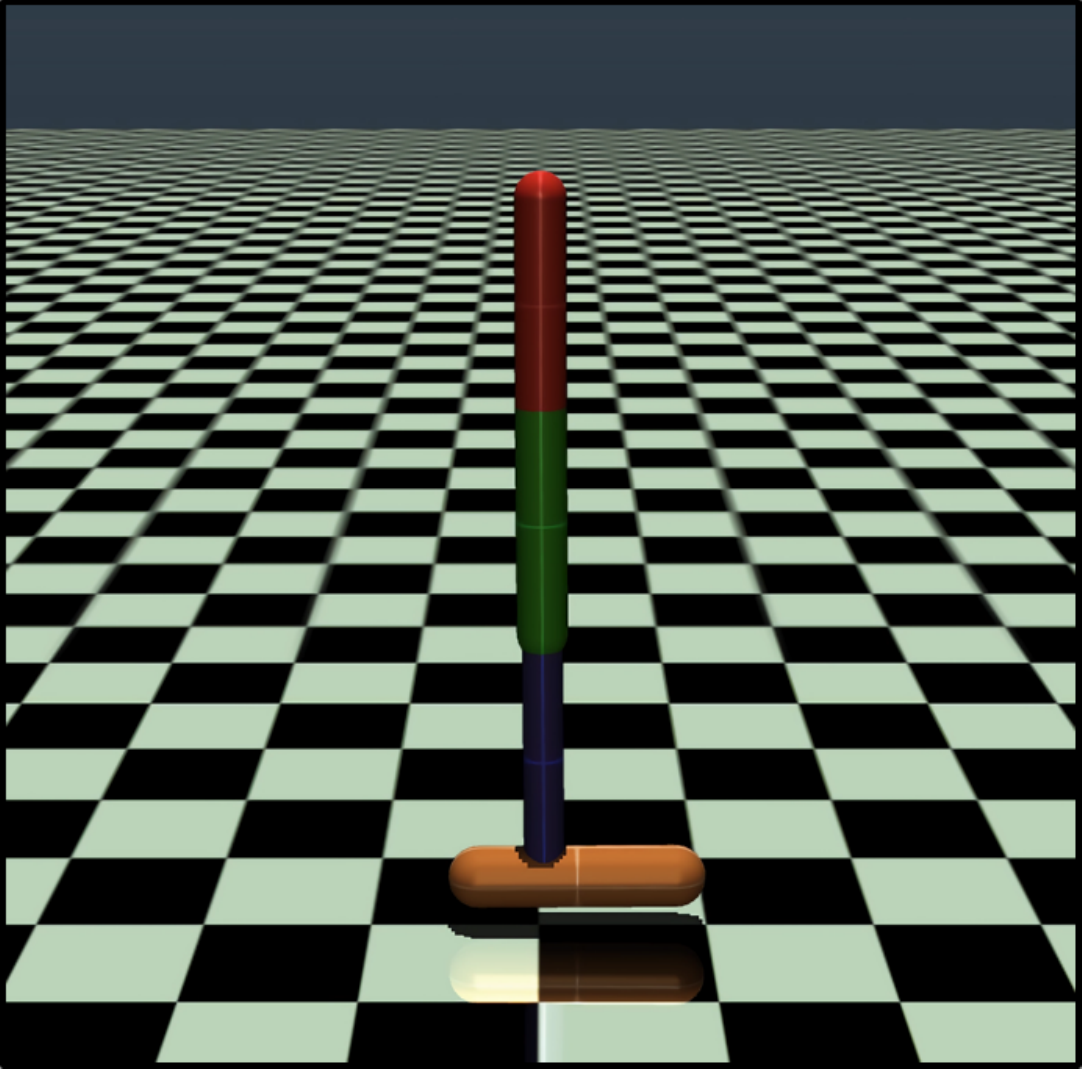}
        \end{minipage}
    }
    \subfloat[HalfCheetah]{
        \begin{minipage}{0.19\linewidth}
            \centering 
            \includegraphics[width=\linewidth]{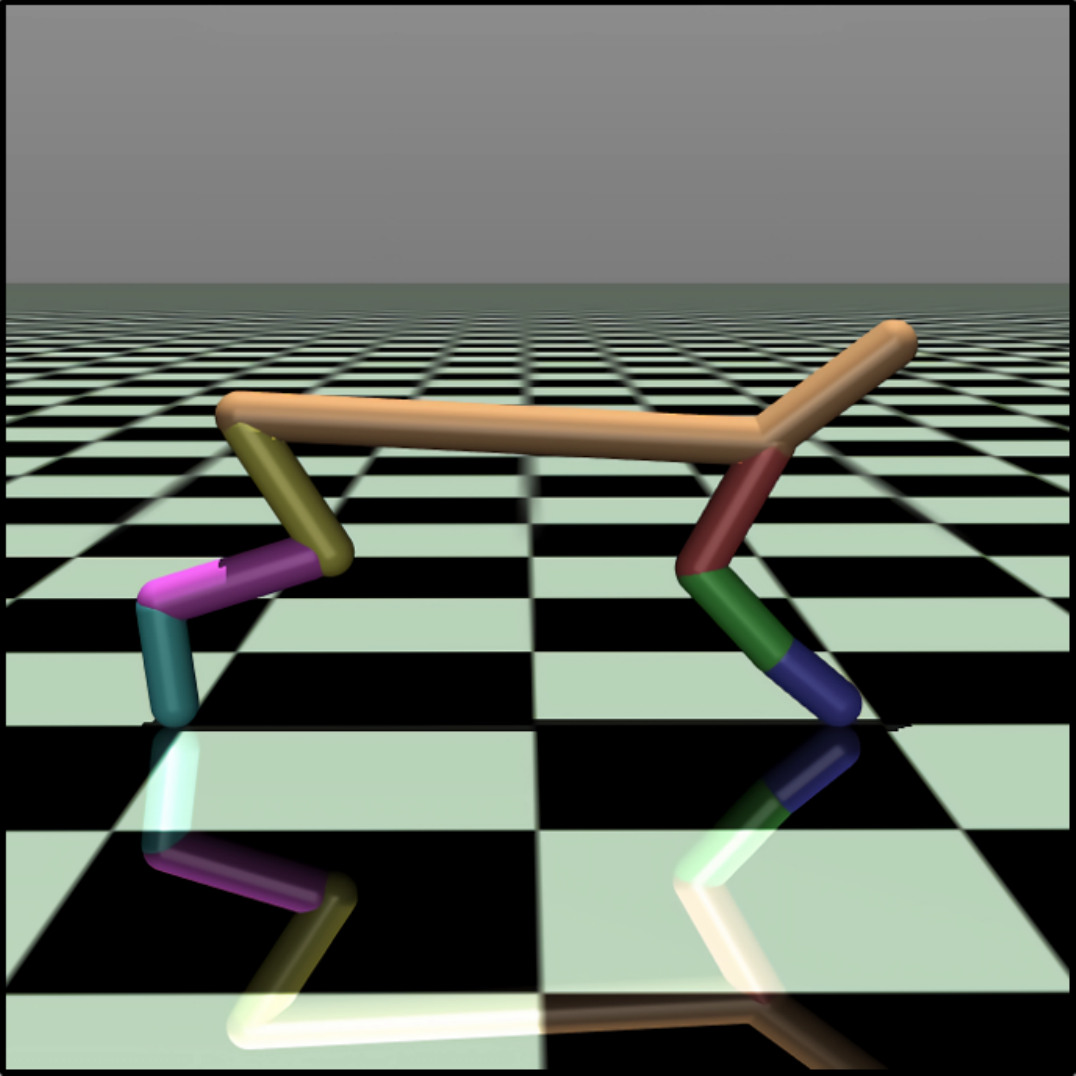}
        \end{minipage}
    }
    \subfloat[Reacher4]{
        \begin{minipage}{0.19\linewidth}
            \centering 
            \includegraphics[width=\linewidth]{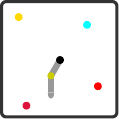}
        \end{minipage}
    }
    \subfloat[Spread]{
        \begin{minipage}{0.19\linewidth}
            \centering
            \includegraphics[width=\linewidth]{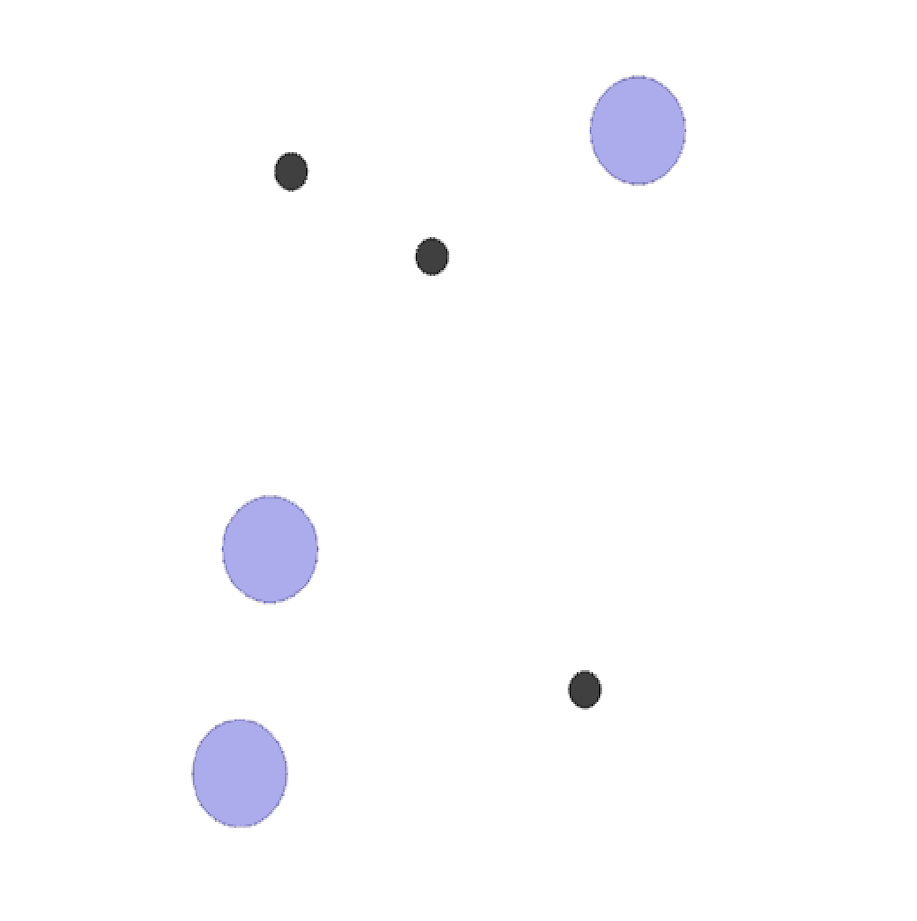}
        \end{minipage}
    }
        \subfloat[Reference]{
        \begin{minipage}{0.19\linewidth}
            \centering 
            \includegraphics[width=\linewidth]{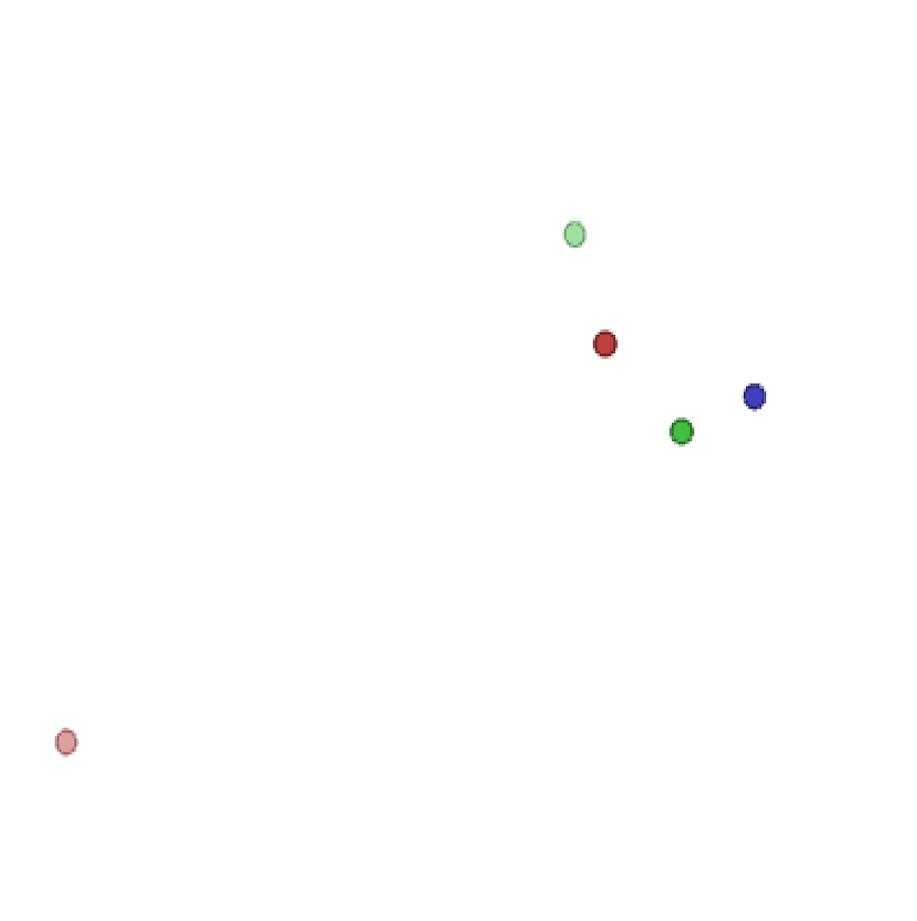}
        \end{minipage}
    }
    \caption{\textbf{Demonstrations of five sparse-reward environments.} Both HalfCheetah (2x3) and HalfCheetah (6x1) are represented as HalfCheetah.}
    \label{fig:env}
\end{figure*}

\begin{itemize}
    \item \textbf{Hopper (3x1)}: This environment features a Hopper in MuJoCo, controlled by agents operating its three primary segments: torso, thigh, and leg.
    Effective forward propulsion requires precise coordination among agents, as the movement of each segment affects the balance and overall progress of the Hopper.
    In sparse-reward settings, agents receive rewards only after the Hopper advances a specified distance, making cooperative exploration essential.
    \item \textbf{HalfCheetah (2x3) and HalfCheetah (6x1)}: In this environment, the HalfCheetah is a robotic model that resembles a simplified, two-dimensional cheetah.
    The two scenarios feature either two agents, each controlling a front or rear leg, or six agents, each controlling a single joint.
    In both cases, the agents must coordinate their actions to achieve forward movement while avoiding falls. Similar to the Hopper (3x1) environment, rewards in sparse-reward settings are granted only when the agent successfully moves forward over a specified distance~\cite{guo2020memory}.
    \item \textbf{Reacher4}: 
    To investigate cooperative exploration in a more complex continuous control setting, we propose a two-agent version of the Reacher environment with four targets, Reacher4.
    The agents are comprised of a two-arm robot, with the goal being to move the robot’s end effector close to one of the four targets.
    In this scenario, we set up sparse rewards, which means that the agents are rewarded only upon making contact with all four targets.
    \item \textbf{Spread}: In Spread, there are $n$ agents and $n$ landmarks.
    The agents' goal is to occupy all landmarks.
    They incur penalties for collisions and are rewarded only when all landmarks are simultaneously covered.
    \item \textbf{Reference}: The Reference scenario requires each agent to approach their designated landmark, which is known only to the other agents.
    Both agents are simultaneous speakers and listeners.
    The key challenge lies in the sparsity of the environment's rewards: agents receive a reward only when all of them reach their designated landmarks simultaneously~\cite{zhang2024mesa}.
\end{itemize}

Across both environments, we consider dense-reward and sparse-reward settings, running $2$ million steps in MPE and $5$ million steps in MAMuJoCo.
In the sparse-reward setting, agents don't receive any intermediate rewards, \textit{i.e.}, agents receive rewards only upon reaching a target or covering a specified distance.

\begin{table*}[t]
  \renewcommand{\arraystretch}{1.2}
  \aboverulesep=0pt
\belowrulesep=0pt
\centering
\caption{\textbf{Performance (higher is better) and time efficiency (lower is better) comparison.} We compare \method, HATD3, and diffusion-based methods in three sparse-reward scenarios, where time denotes the duration of a single iteration.}
\label{tab:time}
\resizebox{\textwidth}{!}{%
\begin{tabular}{>{}lrrrrrrrr<{}}
\toprule
\multicolumn{1}{c}{\multirow{2}{*}{\textbf{Scenarios}}} & \multicolumn{4}{c}{\textbf{Performance} ${\color{violet}\uparrow}$}                                                  & \multicolumn{4}{c}{\textbf{Time (ms)} ${\color{violet}\downarrow}$} \\ \cmidrule(lr){2-5} \cmidrule(lr){6-9}
\multicolumn{1}{c}{} &
  \multicolumn{1}{c}{HATD3} &
  \multicolumn{1}{c}{\method} &
  \multicolumn{1}{c}{\method-DM} &
  \multicolumn{1}{c}{\method-DPM} &
  \multicolumn{1}{c}{HATD3} &
  \multicolumn{1}{c}{\method} &
  \multicolumn{1}{c}{\method-DM} &
  \multicolumn{1}{c}{\method-DPM} \\ \cmidrule(lr){1-9}
Navigation                                     & 8.9$\pm$1.3    & 19.3$\pm$0.3            & \textbf{25.9$\pm$1.8} & 20.5$\pm$1.1  & \textbf{21}    & 48    & 316   & 166   \\
HalfCheetah (2x3)                              & 805.0$\pm$12.6 & \textbf{965.0$\pm$8.0}  & 943.6$\pm$7.2         & 914.3$\pm$7.4 & \textbf{12}    & 37    & 247   & 129   \\
Humanoid (17x1)                                & 902.8$\pm$4.7  & \textbf{1143.6$\pm$9.1} & 991.5$\pm$8.4         & 878.9$\pm$9.5 & \textbf{155}   & 249   & 2133  & 1645  \\ \bottomrule
\end{tabular}%
}
\end{table*}

\paragraph{Baselines}
We compared our results with several baselines as follows.
HATD3 and HASAC are strong continuous control multi-agent algorithms, which are newly extended from TD3 and SAC by HARL~\cite{zhong2023heterogeneousagent}.
MAPPO~\cite{yu2022surprising} stands as a popular and effective MARL algorithm in the continuous domain, generating actions from Gaussian distribution.
MAT~\cite{wen2022multiagent} addresses the multi-agent challenge as a sequential modeling issue, using the Transformer~\cite{vaswani2017attention} to generate complex actions.
Notably, we observe that few MARL methods focus on exploration within continuous action spaces.
Consequently, we extend the classic MARL exploration algorithm MAVEN~\cite{mahajan2019maven} into its continuous version, termed CMAVEN.
Drawing inspiration from FACMAC~\cite{peng2021facmac}, CMAVEN adopts the cross-entropy method (CEM) — a sampling-based, derivative-free heuristic search strategy—for approximate greedy action selection.
CEM has demonstrated effectiveness in identifying near-optimal solutions in nonconvex Q-networks for single-agent robotic control tasks.
In addition, we introduce CMAE~\cite{liu2021cooperative} and adopt the settings described in the original paper, utilizing hash-based counting techniques~\cite{tang2017exploration} to extend the algorithm to continuous state spaces.

\paragraph{Training Details and Hyper-parameters}
We conduct five independent runs with different random seeds in all scenarios.
The experiments in this study were conducted using five distinct seeds on a hardware setup comprising 2 NVIDIA RTX A6000 GPUs and 1 AMD EPYC CPU.
We utilized official implementations of baseline algorithms, adhering strictly to their original hyper-parameters.
The employed consistency policy involves a multi-layer perceptron (MLP), which processes state inputs to generate corresponding actions.
Specifically, our policy networks incorporate a 3-layer MLP architecture with a hidden size of 256, using the Mish activation function.
The intention learner consists of a MLP, an intention codebook module, and another MLP.
The MLP network is a 128-dimensional fully-connected layer, followed by a GELU activation, and followed by another 128-dimensional fully-connected layer.

For the consistency policy, we define the diffusion step $k$ within the range of $[0.002,80.0]$, setting the number of sub-intervals $M=40$.
Following Karras diffusion model, the sub-interval boundaries are determined with the formula $k_{i}=\left(\epsilon^{\frac{1}{\rho}}+\frac{i-1}{M-1}\left(T^{\frac{1}{\rho}}-\epsilon^{\frac{1}{\rho}}\right)\right)^{\rho}$, where $\rho=7$.
The Euler method is employed as the ODE solver.
In the intention learner, we set the number of discrete intentions in codebook $K=5$ in different scenarios, and $\beta=0.2$.
The performance of intention learning is not sensitive to the setting of $K$.
Regarding performance evaluation, we pause training every $M$ steps and evaluate for $N$ episodes with action selection. 
The $(M,N)$ in MPE and MAMuJoCo are $(1000,20)$ and $(10000,40)$, resepcetively.
The training objective is optimized by Adam with a learning rate of $5\times 10^{-4}$.

\subsection{Performance Comparison}
\label{sec:results}
We first compare \method with baselines across six scenarios.
\cref{tab:comparison} outlines the comparison between our approach and various baseline algorithms in both dense reward and sparse reward settings.
In \emph{dense-reward} environments, \method demonstrates performance slightly below or comparable to baselines, surpassing them in specific tasks.
In the context of online MARL in POMDPs, the theoretically optimal policy tends to be deterministic, suggesting that expressive models often revert to unimodal to approach optimality. 
Consequently, expressiveness primarily aids in exploration rather than contributes to the final optimal policy convergence.
In \emph{sparse-reward} environments requiring exploration, \method approach exhibits a significant performance advantage, outperforming baseline algorithms by 20\%.
While methods like HATD3, MAPPO, and MAT show proficiency in dense reward contexts, their limited focus on exploration hinders their effectiveness in reaching sparse reward states.
Although the HASAC method improves exploration capability through entropy regularization, the disparity between its employed Gaussian policy and the assumed theoretical Boltzmann policy leads to local optima, thereby constraining exploration.
Notably, despite being tailored for MARL exploration, CMAVEN and CMAE struggle considerably in continuous action or state spaces, resulting in underwhelming performance.
In contrast, our proposed method, employing the consistency policy with intention guidance and a self-inference mechanism, fosters more efficient exploration and significantly enhances performance.

\subsection{Time Efficiency}
\label{sec:time}
Alongside its superior performance, we observe that \method also possesses higher time efficiency compared with other methods based on diffusion models.
This efficiency is essential for practical scalability in multi-agent systems and for facilitating real-time decision-making processes.
To demonstrate this, we replace the policy class in our method with a diffusion model (\method-DM) and an accelerated variant of it (\method-DPM)~\cite{lu2022dpmsolver}, and compared the time required by the different algorithms for each training iteration, as shown in \cref{tab:time}.

In \texttt{Navigation} and \texttt{HalfCheetah(2x3)} scenarios, our method attains training speeds up to $6.6\times$ and $6.8\times$ faster than those of diffusion-based methods, rivaling the time efficiency of previous methods like HATD3.
Despite using DPM-Solver, \method-DPM still faces challenges in achieving satisfactory training efficiency.
Moreover, the training speed of algorithms is greatly impacted by the scaling in MAS.
We selected \texttt{Humanoid(17x1)} scenario, which involves 17 agents, for evaluating our methods under large-scale agents.
In this scenario, the training time of diffusion-based methods significantly exceeds that of \method, demonstrating the scalability of our method in MARL.

Due to the substantial reduction in sampling steps, the consistency model is inherently less expressive than the diffusion model~\cite{song2023consistency}.
To evaluate this, we conduct a performance comparison between \method and diffusion-based methods in three sparse scenarios.
In comparison to \method-DM and \method-DPM, \method exhibits a performance decrease in MPE scenarios, yet outperforms it in the \texttt{HalfCheetah(2x3)} and \texttt{Humanoid (17x1)} environments.
This phenomenon could potentially be attributed to that the consistency policy can effectively transmit action gradients in one step, thereby alleviating the challenges of policy training via Q-function in diffusion-based methods~\cite{chen2023boosting}.
Moreover, the slight performance trade-off in \method is considered acceptable, given its significant improvements in training and inference efficiency.

\subsection{Qualitative Analysis and Visualization}
\label{sec:visual}

\begin{figure}[tb]
	\centering
	\includegraphics[width=\linewidth]{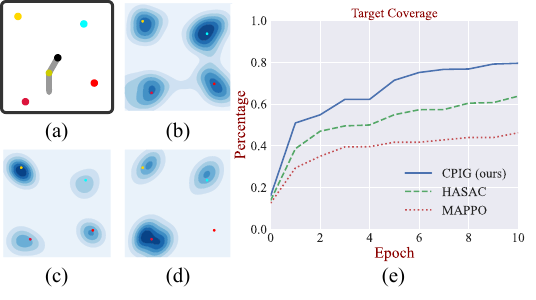}
	\caption{\textbf{Example results obtained by \method and baselines on exploration visitation.}
		(a) shows a Reacher4 task, with four targets in different colors.
		(b)-(d) shows the state visitation of \method, HASAC and MAPPO.
		(e) counts the targets covered by policies during exploration.
	}
	\label{fig:reacher}
\end{figure}

To showcase the exploration capabilities of our method, consider a didactic example illustrated in \cref{fig:reacher}(a), which demonstrates different MARL algorithms applied to the sparse \texttt{Reacher4} task \cite{peng2021facmac} featuring four targets.
The agents consist of two-arm robots tasked with moving the robot's end effector close to one of the four targets.
The depth of color represents the magnitude of the state visitation probability.
In this scenario, we configure sparse rewards, meaning that the agents receive rewards only upon making contact with all four targets.
Besides, the agents need to cooperate in exploration to reach the target, and the dispersion of targets often leads to agents adopting one single policy, underscoring the necessity for multimodal policies and cooperative exploration.
In \cref{fig:reacher}(b), our method learns a multimodal joint policy, enabling the effector to reach four different targets relatively uniformly, without converging to a unimodal policy.
Conversely, in \cref{fig:reacher}(c) and (d), baselines primarily focus on visiting certain targets, thereby limiting exploration in the complex environment.
Additionally, we quantify the targets covered by the policy during exploration in \cref{fig:reacher}(e).
The curves indicate that our method achieves a state visitation rate of 83\%, compared to 62\% for HASAC, reflecting a 20\% improvement.
These results demonstrate that \method can facilitate a more efficient exploration of the four different targets in the scenario, highlighting the importance of multimodal policies and cooperative exploration capabilities among agents.

\begin{figure}[tb]
	\centering
	\includegraphics[width=\linewidth]{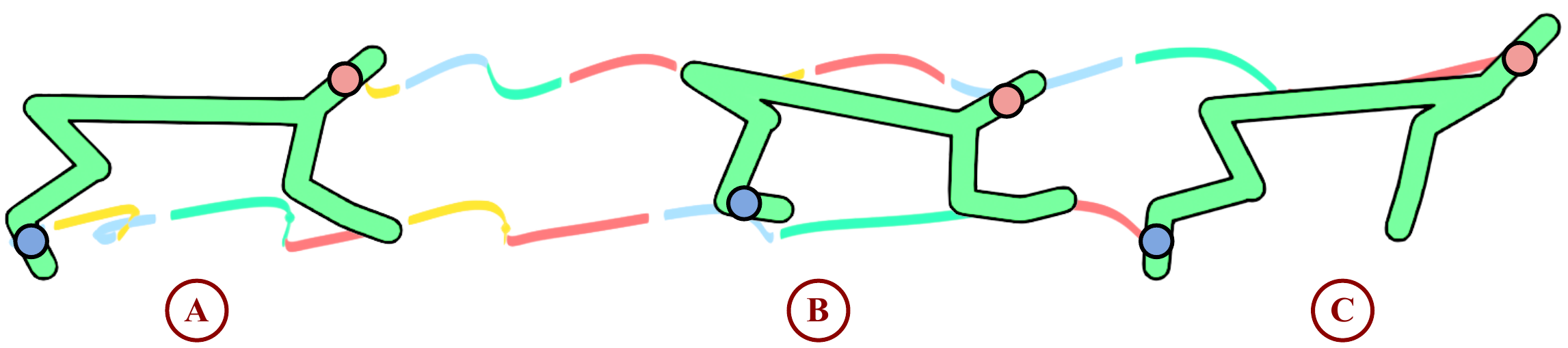}
	\caption{\textbf{Trajectory visualization in HalfCheetah (2x3).}
		Different colors represent the corresponding intention between agents.
	}
	\label{fig:mujo_task}
\end{figure}

\begin{figure}[tb]
	\centering
	\subfloat[Navigation]{
		\begin{minipage}{0.47\linewidth}
			\centering
			\includegraphics[width=\linewidth]{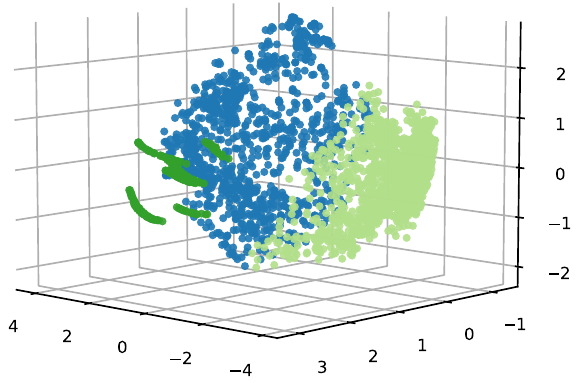}
		\end{minipage}
	}
	\subfloat[HalfCheetah (2x3)]{
		\begin{minipage}{0.47\linewidth}
			\centering 
			\includegraphics[width=\linewidth]{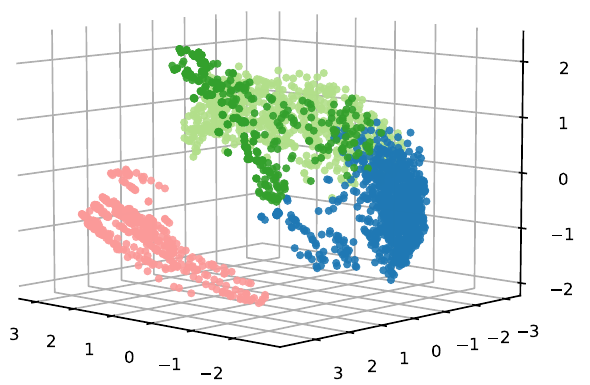}
		\end{minipage}
	}
	\caption{\textbf{Visualization of the observation embeddings.} We choose the last $30k$ steps for visualization. The colors denote different intentions.}
	\label{fig:intention_visal}
\end{figure}

In order to gain a better understanding of the intention learner within \method, we generate visualizations of intention guidance and observation embeddings.
In \cref{fig:mujo_task}, we visualize the \emph{true} trajectories generated by the intention-guided consistency policy in the \texttt{HalfCheetah(2x3)} scenario.
The trajectories are color-coded to represent the corresponding intention generated during execution.
Across stages \textbf{A}, \textbf{B}, and \textbf{C}, although the observations of the two agents differ, they converge to a common intention (the same color).
This indicates that the intention effectively reflects the global state, enabling multiple agents to coordinate their actions.
By leveraging this shared intention, multiple agents can cooperate more effectively, thereby enhancing their exploration of the environment.
Additionally, we present the visualizations obtained by applying the PCA
technique to the observation embeddings.
The goal is to show whether the intention learner can deduce distinguishable intention from the local observations, aiding in the guidance of policy during the exploration process.
As shown in \cref{fig:intention_visal}, the visualizations demonstrate that the intention learner can generate informative discrete representations by identifying potential correlations within individual observations of agents.

\subsection{Ablation Study}
\label{sec:abla}

\begin{figure}[t]
	\centering
	\includegraphics[width=\linewidth]{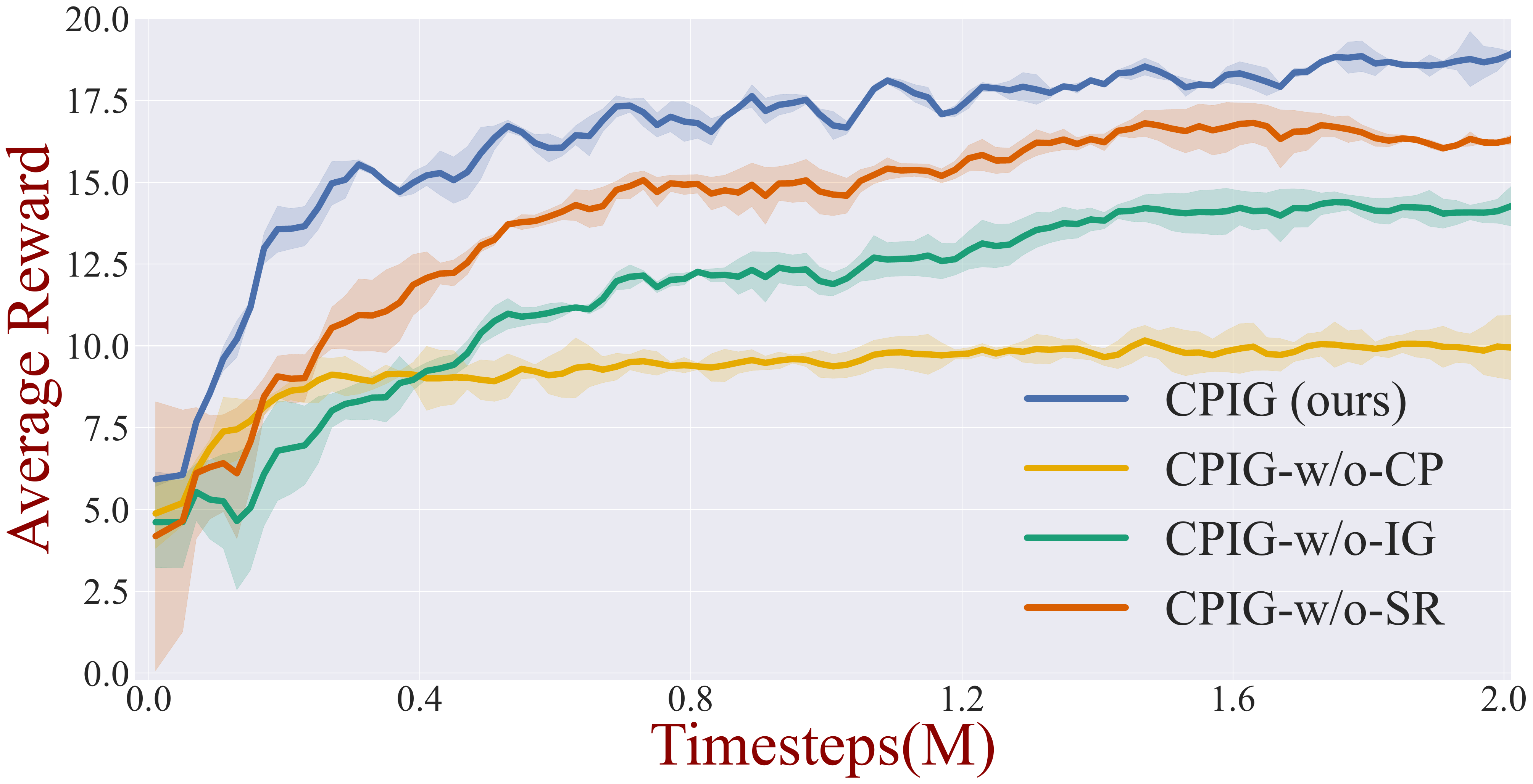}
	\caption{The ablation performance of the three key components in our method under the sparse Navigation scenario.}
	\label{fig:abla}
\end{figure}

In this subsection, we conduct ablation studies in the \texttt{Navigation} scenario to investigate the impact of three main components in \method:
(1) replacing the consistency policy with a deterministic policy, similar to MADDPG~\cite{lowe2017multi}, while directly incorporating intention guidance and observations directly as the inputs (\emph{\method-w/o-CP}),
(2) removing the intention learner module, forcing agents to rely solely on their own observations for exploration during training (unguided exploration) and for achieving objectives during execution (\emph{\method-w/o-IG}), and
(3) eliminating the self-reference mechanism, which relaxes the constraints on consistency policies (\emph{\method-w/o-SR}).
The experimental results, as shown in \cref{fig:abla}, demonstrate that \emph{\method-w/o-CP} quickly converges to a locally optimal policy, highlighting the importance of the consistency policy in enabling effective exploration and preventing premature convergence to suboptimal solutions. 
Additionally, \method achieves superior performance compared to \emph{\method-w/o-IG}, which lacks intention guidance, as evidenced by its faster convergence rate and higher final performance.
Similarly, when using only intention guidance, \emph{\method-w/o-SR} exhibits a lack of exploitation of past successes, reducing sample efficiency and leading to performance degradation.
Overall, these findings underscore the importance of all three components—consistency policy, intention learner, and self-reference mechanism—in driving the improved performance of \method.

\section{Conclusions and Future Work}
In this paper, we propose \method, a novel multi-agent exploration method that incorporates an intention-guided consistency policy to enable cooperative exploration and a self-reference mechanism to constrain generated actions.
To the best of our knowledge, this study represents the first effort to leverage consistency policies in the context of MARL.
Empirical results from six environments demonstrate that \method achieves performance comparable to baseline algorithms in dense reward settings while in sparse reward settings, which pose more challenges for exploration, our algorithm outperforms baselines by 20\%.

This work is the first to showcase the benefits of intention-guided consistency policies in the realm of multi-agent exploration.
However, the present work is confined to exploration within single-task scenarios.
In the future, we intend to explore how to better harness the multimodal nature of consistency policies to devise a faster and more efficient algorithm for multi-task MARL.

\bibliographystyle{IEEEtran}

\bibliography{IEEEabrv,ref}

\end{document}